\documentclass[journal, 12pt, onecolumn]{IEEEtran}
\usepackage{amsfonts}
\usepackage{amssymb}
\usepackage{graphicx}
\usepackage{amsmath}
\usepackage{multirow}
\usepackage{epsfig,graphics}
\usepackage{cite}
\usepackage{color}
\usepackage{makecell}
\usepackage[caption=false,font=footnotesize,subrefformat=parens]{subfig}	

\begin{document}
\baselineskip 24pt
\parskip 9pt
\title{Internet of Things Cloud: Architecture and Implementation}

\author{Lu Hou$^\ast$, Shaohang Zhao$^\ast$, Xiong Xiong$^\ast$, Kan Zheng$^\ast$,~\IEEEmembership{Senior Member,~IEEE}, Periklis Chatzimisios$^\dagger$,~\IEEEmembership{Senior Member,~IEEE}, M. Shamim Hossain$^\S$, Wei Xiang$^\ddagger$,~\IEEEmembership{\it Senior Member,~IEEE}\\

\thanks{
$^\ast$Lu Hou, Shaohang Zhao, Xiong Xiong and Kan Zheng are with the Intelligent Computing and Communication (IC$^2$) Lab, Wireless Signal Processing and Networks Lab (WSPN), Key Lab of Universal Wireless Communications, Ministry of Education, Beijing University of Posts and Telecommunications, Beijing, China, 100088. (e-mail: zkan@bupt.edu.cn).
}

\thanks{
$^\dagger$Periklis Chatzimisios is with the CCSN Research Lab, Alexander Technological Educational Institute of Thessaloniki (ATEITHE), 57400 Sindos, Thessaloniki, Greece.
}

\thanks{
$^\S$M. Shamim Hossain is with the Software Engineering Department, College of Computer and Information Sciences, King Saud University, Riyadh 11543, KSA.
}

\thanks{
$^\ddagger$Wei Xiang is with the College of Science and Engineering, James Cook University, Cairns, QLD 4878, Australia.
}
}
\maketitle

\begin{abstract}
The Internet of Things (IoT), which enables common objects to be intelligent and interactive, is considered the next evolution of the Internet. Its pervasiveness and abilities to collect and analyze data which can be converted into information have motivated a plethora of IoT applications. For the successful deployment and management of these applications, cloud computing techniques are indispensable since they provide high computational capabilities as well as large storage capacity. This paper aims at providing insights about the architecture, implementation and performance of the IoT cloud. Several potential application scenarios of IoT cloud are studied, and an architecture is discussed regarding the functionality of each component. Moreover, the implementation details of the IoT cloud are presented along with the services that it offers. The main contributions of this paper lie in the combination of the Hypertext Transfer Protocol (HTTP) and Message Queuing Telemetry Transport (MQTT) servers to offer IoT services in the architecture of the IoT cloud with various techniques to guarantee high performance. Finally, experimental results are given in order to demonstrate the service capabilities of the IoT cloud under certain conditions.
\par
\end{abstract}

\begin{flushleft}
\textbf{\textit{Index Terms}}-- Internet of Things (IoT), Cloud Computing, IoT applications, IoT architecture
\end{flushleft}

\baselineskip 24pt
\parskip 9pt

\section{Introduction}
\label{sec_intro}

With the development of wireless communication technologies, pervasive objects can be interactive and are connected to the Internet. These inter-connected objects with in-built computing, communications and sensing capabilities constitute the Internet of Things (IoT). In particular, it is estimated that by 2020 the number of IoT devices will be close to 50 billion while the population will reach 7.6 billion~\cite{WP_IoT}. These devices can generate huge amounts of data, which usually come in different format and meaning~\cite{bigdata},~\cite{data},~\cite{bayesian}. However, IoT devices usually have very limited capabilities due to their small physical size and energy consumption. Therefore, an IoT cloud is imperative to support the requirements of millions of IoT devices and provide various new and exciting IoT applications for the end-users. 
\par

Recently, a plethora of novel ideas has been proposed for the IoT cloud. Machine-to-Machine (M2M) communications or Machine Type Communications (MTC), which enable direct communication among IoT devices, have attracted significant attentions. A standard M2M service layer platform, which is called oneM2M, has been established and developed for the standardization of deployment of IoT services~\cite{oneM2M}. A detailed discussion over MTC can be found in~\cite{mtc}. Subscription control, congestion and overload control are described in detail, as well as a new solution to the latter issue. As for the IoT itself, the authors in~\cite{iot0} and~\cite{enabling} provide a comprehensive survey and discuss the feasibility as well as enabling technologies for the IoT cloud. Meanwhile, inspired by the IoT cloud, certain research on related applications is conducted in~\cite{DoS},~\cite{D2D},~\cite{Service}. Moreover, in order to efficiently manage IoT workloads, several IoT cloud platforms have been proposed. For example, the authors in~\cite{servIoTicy} propose a platform dubbed servIoTicy with data stream processing capabilities for the IoT cloud and carry out a performance evaluation of the proposed platform. However, the relevant literature on the architectural design and implementation details for the IoT cloud are scarce to date. An architecture that can support millions of concurrent IoT devices as well as diverse IoT applications is highly desirable.
\par

Against the above background, this paper aims at proposing an IoT cloud architecture based on both the Hypertext Transfer Protocol (HTTP) and Message Queuing Telemetry Transport (MQTT) protocols, and other relevant techniques to guarantee high performance. Firstly, certain application scenarios and requirements of the IoT cloud are presented. A generic IoT system is then proposed and we discuss the supporting IoT infrastructure. Moreover, the implementation details of the proposed IoT cloud architecture are discussed, followed by a presentation of the way that the IoT system is build and how a message broker for MQTT server is built by employing the Redis cluster database. Finally, a number of experiments is conducted to evaluate the performance of the IoT cloud, in terms of the average response time/average transmission latency, throughput and CPU utilization.
\par

The rest of paper is organized as follows. Section II provides some typical application scenarios for the IoT cloud and the corresponding performance requirements. The proposed IoT cloud architecture is then described in Section III. Section IV discusses in detail the overall implementation of the proposed IoT cloud, while Section V reports on our experimental results. Finally, Section VI concludes this paper and provides a future outlook.
\par

\section{Typical Applications of the IoT Cloud}

The use of an IoT cloud has a significant impact on the performance of the supported IoT applications. Thus, several typical application scenarios of the IoT cloud are discussed in this Section, including:
\par

\subsubsection{Smart Buildings}
Smart buildings can adapt to internal and external environmental changes without human intervention in order to provide comfort to the occupants, while taking into consideration financial and energy requirements. Ubiquitous devices can monitor the entire building at all times, generating large amounts of data, which can then offer notification services in case of emergency incidents or critical situations after a proper analysis on the gathered data~\cite{city}.
\par

\subsubsection{Smart Home/Office}
A smart home establishes a future home environment, where embedded sensors and intelligent appliances are self-configured and can be controlled remotely through the Internet. It enables a variety of monitoring and intelligent control applications that are responsible for the control and management of home resources. More intelligently, an IoT device can be controlled by the IoT cloud to adjust its operation. As a result, a comfortable living environment can be created for everybody. On the other hand, smart offices aim mainly at easing the workload and improving the productivity of employees at work. With a proper IoT cloud, workers in different organizations or areas can all access to office-related services in a convenient and efficient manner.
\par

\subsubsection{Intelligent Transportation}

Intelligent transportation brings comfort to people in travel within cities or rural areas. Vehicles can be smart if they are equipped with a large number of sensors that monitor the status of the surrounding environment. All sensed data can be collected and uploaded to the IoT cloud. With real-time data processing, the IoT cloud can provide useful assistance to the driver such as the provision of emergency warnings or optimal path planning, as well as the knowledge of road/traffic conditions or traffic accident notifications. The IoT cloud also offers warnings for pedestrians when there are under potential life or injury threats by analyzing the data gathered from vehicles, infrastructure or even pedestrians.
\par

\subsubsection{Smart Healthcare}

Smart healthcare applications decrease patients’ dependence on carers and reduce their healthcare costs through efficient use of medical equipment and sensors. The measurement of various biological information such as the pulse and blood pressure can be taken by the patients themselves via their smartphones that contain special on-body and near-body sensors. All the measured data are then transmitted to the IoT cloud, where the early detection of life-threatening emergency situations can be made possible through continuous monitoring and analyzing the received data from the patient under monitoring.

\par

For the purpose of better illustration, we summarize the features of several typical application scenarios using the IoT cloud in Table~\ref{table_application}.
\par

\section{Architecture of the IoT systems}

In order to support millions of IoT devices, we propose an IoT cloud architecture based on the hardware support of the IoT infrastructure. By using virtualization, hardware resources can be well utilized. Consequently, both HTTP and MQTT servers are introduced as the application servers of the IoT cloud. The HTTP servers can provide services for end-users and devices, while the MQTT servers ensure a large number of device connections and real-time communication among devices. Furthermore, some other key components such as the supporting databases are also presented for the sake of functionality, availability and performance.

\subsection{IoT infrastructure}
IoT infrastructure is a fundamental component of the entire IoT system since it can sense and perform actions from/to the environment as well as sending information to the IoT cloud. The IoT infrastructure consists of all IoT devices and the supporting access networks. The former is deployed in the application environment, whereas the latter provides communications between IoT devices and the cloud. IoT devices mainly include sensors, actuators, intelligent appliances, etc. and may generate huge amounts of data that are transmitted to the IoT cloud through reliable and efficient access networks. Additionally, control messages may be transferred to IoT devices from the IoT cloud via the same access networks.
\par

\subsection{IoT Cloud}
\begin{figure}[!t]
	\centering
	\includegraphics[width=6.5in]{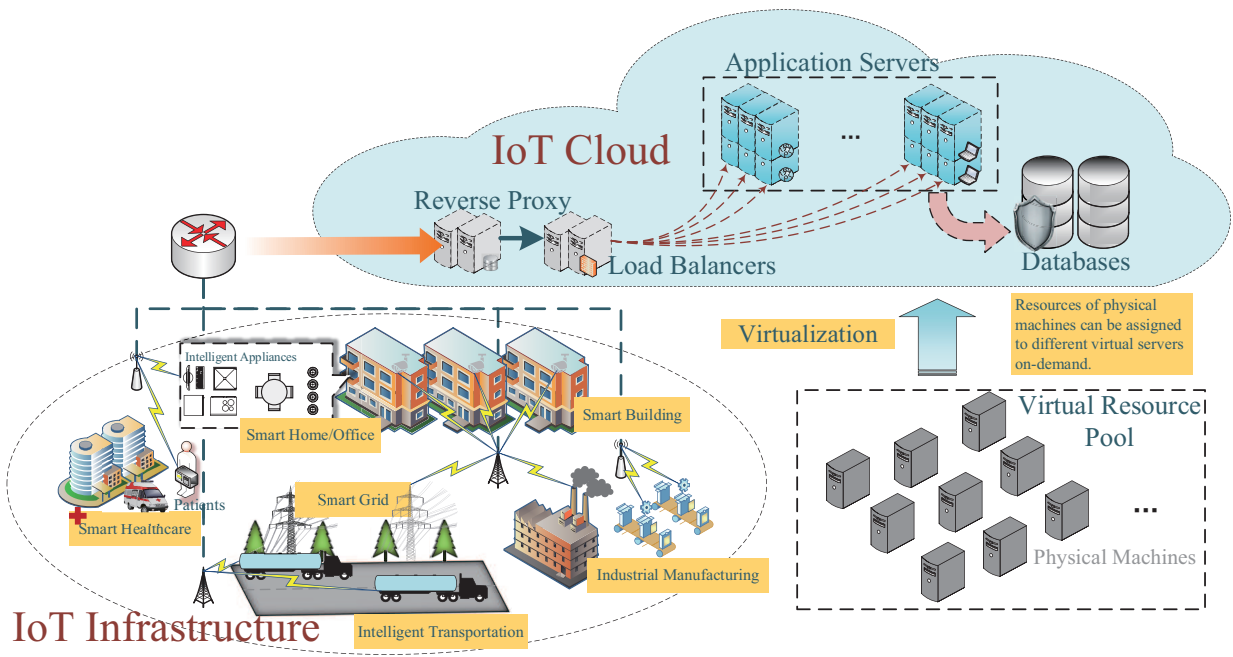}
	\caption{Illustration of the architecture of IoT systems}
	\label{fig_architecture}
\end{figure}

As illustrated in Fig.~\ref{fig_architecture}, the IoT cloud consists of several key components, each of which is composed of multiple servers that perform different tasks. The servers are established as Virtual Machines (VMs) utilizing virtualization technology. They are independent from each other even if they run on the same physical machine. With these VMs, load balancers/reverse proxy servers, databases and application servers can be configured. The functionalities of each component are described as follows, i.e.,
\par

\subsubsection{Virtual resource pool}
As the hardware resources of the physical machine (such as the CPU, memory and network connectivity) cannot be fully utilized, there is a significant waste of resources, as well as the problem of low scalability of servers. To tackle these problems, the virtualization technique is used to provide feasible solutions that aim at improving resource utilization for the IoT cloud. By means of virtualization, hypervisor software runs on the physical machine as an abstract layer to manage all resources and also to provide an operating environment for various independent guest Operating Systems (OSs) (known as VMs) that enable dynamic resource allocation. Furthermore, the IoT cloud services can be deployed on VMs instead of directly on physical machines, which helps reduce the usage of physical machines and, thus, can deliver high performance at low cost.
\par
In particular, through employing the virtualization technique, a virtual resource pool can be established on several physical machines that contain all the hardware resources and can assign them to different VMs on demand. In this way, all other servers can obtain a proper amount of resources, in accordance with their demands. 
\par

\subsubsection{Application servers}
Application servers are often considered as the most important component of the IoT cloud since they are responsible for offering business services to customers. They need to provide facilities and an appropriate environment to run multiple applications based on certain application protocols~\cite{protocol}. The application servers in the traditional cloud are usually based on the HTTP. HTTP servers work in a request-response manner through the Transmission Control Protocol (TCP) connections with clients. When connections are established, an HTTP server can listen to certain ports for requests from clients and send appropriate responses to the received requests.
\par

However, HTTP is not well suited for the IoT cloud since IoT devices are constrained by their computing, communication, and energy resources. Consequently, another type of application protocol is more appealing for the IoT cloud, i.e., the MQTT protocol~\cite{protocol}. MQTT is designed for resource-constrained IoT devices as a lightweight messaging transportation protocol that operates via a topic-based publish-subscribe mode. This means that when a client publishes a message on a particular topic, all the clients that have subscribed to the same topic can receive this message. A key component that completes the transfer process is regarded as the broker~\cite{broker}, by which one-to-many connections are enabled.
\par

\subsubsection{Database}
According to various application requirements for data storage, relational and non-relational databases, also known as Structured Query Language (SQL) and NoSQL databases, are optional in the IoT cloud. SQL is designed as a type of programming language for relational databases that can store data in the form of two-dimensional tables. However, the performance of the SQL databases is the main bottleneck for the deployment of real-time IoT applications. Consequently, NoSQL databases are used to provide real-time and high-efficiency services for data storage. These databases allow data to be stored directly in memory or hard disks and, thus, the Input/Output (I/O) speed is significantly improved.
\par
 
\subsubsection{Reverse Proxy and Load balancing}
Due to the large number of IoT devices and users, application servers are required to handle millions of concurrent requests or transfer massive number of messages. These requests or messages are processed without scheduling if load balancing is not enforced. As a result, some servers may be heavily congested due to excessive burdens, and it is possible that new requests or messages sent to the congested servers are rejected or discarded. Meanwhile, other servers may be idle, although spare resources may actually be enough to process these requests, resulting in a significant waste of resources. Therefore, load balancing is imperative for the even distribution of workload across multiple backend servers and achieve full utilization of all available resources.
\par

\section{Implementation and Services of the IoT Cloud}
Aiming at connecting millions of devices and end-users, the proposed IoT cloud is developed with details that are elaborated in this Section. The services that the IoT cloud can provide are also discussed.
\par

\subsection{Implementation}
A virtualization OS such as VMWare vSphere\footnote{http://www.vmware.com/products/vsphere/} can be used to establish a resource pool with a number of VMs and can directly handle the CPU and memory resources of the physical machines. A server in the IoT cloud can be implemented as one VM. The implementation of different kind of servers are described in detail as follows, i.e.,

\subsubsection{Application servers}
The IoT cloud includes the HTTP and MQTT servers that both can be developed using Node.js, which is typically used for developing server applications due to its capability in high concurrency. It runs in the form of an asynchronous event loop which performs all I/O operations with a single thread asynchronously. As a result, application servers are capable of handling a large number of concurrent connections.
\par

\paragraph{HTTP servers}
They apply a flexible web application framework, i.e., Express, in order to work. In such a way, the web and mobile applications are easily deployed on an HTTP server, which interacts with clients through a Request-Response cycle. The HTTP servers offer three different methods, i.e., GET, POST and DELETE for clients to make requests. Clients can obtain resources from the HTTP servers through a GET request. Clients can also send information to the HTTP servers through a POST request. Moreover, a DELETE request enables clients to delete certain resources in an HTTP server. After receiving a request, the HTTP server tries to process the request and send a response back to the clients.
\par

\paragraph{MQTT servers}
They are deployed for instant communication between the IoT devices and end-users by utilizing the MQTT protocol that is a broker-based protocol for publishing/subscribing message transportation. Its publication and subscription are organized based on the notion of ``topic'' and all packets are published through the broker. As for publication, a topic should be uniquely defined, while for subscription, an MQTT client can subscribe multiple topics at once. The MQTT servers are implemented based on an open-source library in Node.js, i.e., MQTT-connection. In order to enhance real-time performance of the MQTT servers, they have to maintain long-lived TCP connections with clients or devices. Furthermore, the MQTT servers use three levels of Quality of Service (QoS) to ensure reliability. QoS level 0 means that recipients do not send any acknowledgment to publishers, and all messages are published only once. By contrast, QoS level 1 requires acknowledgments. As for QoS level 2, a handshake mechanism is used to make sure that messages can be successfully delivered to all subscribers. In accordance with the various business requirements, the corresponding QoS level can be configured.
\par

Multiple application servers can constitute a cluster, which allows for simultaneously scaling server programs across multiple parallel processors. The Parallel Multithreaded Machine (PM2) helps form a cluster of multiple HTTP servers. On the other hand, the MQTT server has to act as a cluster through operating in a master-slave mode. In this mode, the MQTT server operates at the master node, which can initiate other servers as the slave nodes. Each MQTT server (master or slave) runs on an individual CPU core.
\par

\subsubsection{Database cluster and broker}
\begin{figure}[!t]
	\centering
	\includegraphics[width=6.5in]{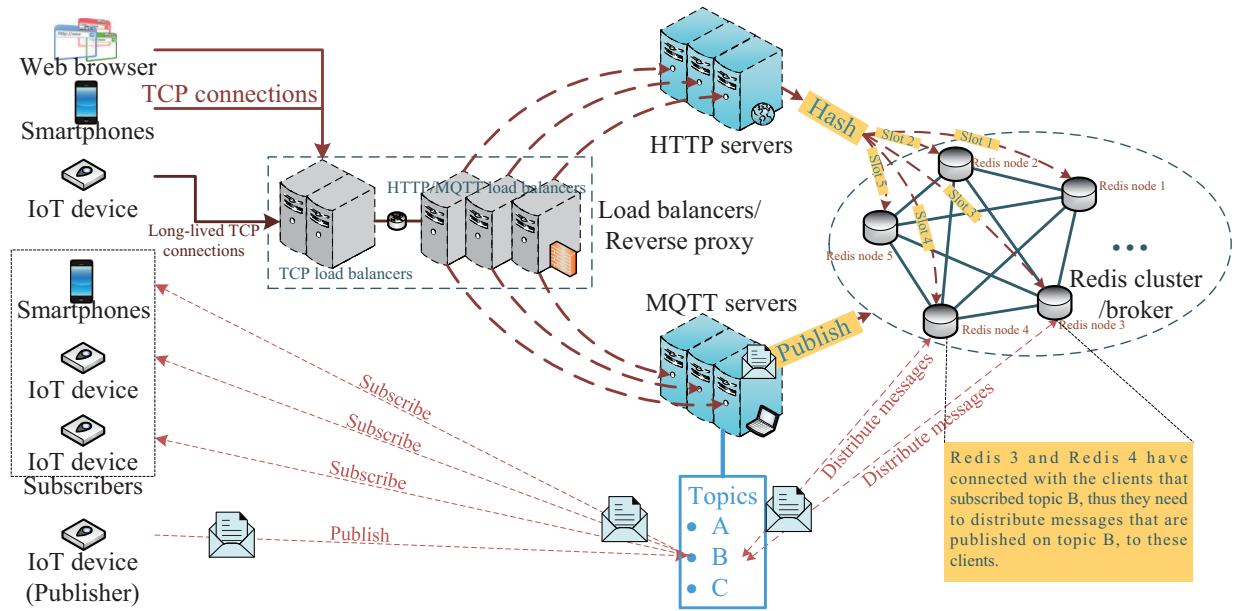}
	\caption{Illustration of the IoT cloud implementation}
	\label{fig_implementation}
\end{figure}

The IoT cloud uses Redis\footnote{http://redis.io/} (a NoSQL database) to store data. By storing all key-value data in the memory, Redis can significantly increase the I/O speed. In order to improve the reliability of the database, a Redis cluster with more than one Redis nodes can be configured in the IoT cloud. Thus, the users can enjoy continuous data services even when one or more Redis nodes are out of order. The Redis cluster is fully connected such that each Redis node is connected with all the others through TCP connections. After forming the Redis cluster, slot share should be configured before the cluster can work properly. The data stored in the Redis cluster are firstly hashed, e.g., taking the CRC16 of modulo 16384 of the data as the hash slot. By checking in which interval the hash slot is located, the data are then stored in the corresponding Redis node, as depicted in Fig.~\ref{fig_implementation}. Since the hash slots of incoming data are uniformly distributed, the load of each Redis node is inherently balanced. From the users' viewpoint, there is no difference in accessing the database whether it is a single Redis node or a cluster.
\par

On the other hand, Redis works well with the MQTT servers, since it can work as a message broker. The load of the MQTT server can then be largely moved to the Redis cluster so that higher concurrency can be achieved. Clients can publish messages on some topics to the MQTT servers at first, which then transfers the payloads of the messages directly to the Redis cluster. These payloads can only be received by one Redis node. However, this node may not be connected with the clients that subscribe to these topics. Therefore, this Redis node would have to share the payloads with all the other nodes. Other nodes connected with the correct subscribers can send the payloads to those subscribers. In this manner, a satisfactory performance of the message broker can be ensured.
\par

To further guarantee database reliability, advanced techniques such as hot standby and transaction logging are needed for the Redis cluster. Each Redis node has its backup, which runs in a standby server in the event of malfunction. Furthermore, the transaction logging can record the history of the database including hostile attacks for the convenience of recovery.
\par

\subsubsection{Load balancers}
In order to support a large numbers of requests and messages from massive IoT devices and users, a load balancer is necessary for the IoT cloud. HAProxy\footnote{https://www.haproxy.org/} is a proper option. As an open-source light-weight load balancer, HAProxy can offer efficient TCP-based and HTTP-based load balancing for the IoT cloud and runs in an event-driven, single-thread model supporting high concurrency. Through proper configurations, two types of load balancing are deployed for the IoT cloud with HAProxy, i.e.,
\par

\paragraph{HTTP load balancers}
The default load balancing method, i.e., Weighted Round Robin, is configured in HAProxy to distribute requests for the HTTP servers. HAProxy can check the Uniform Resource Locator (URL) of a request on a binding port, and then distributes the request to the object HTTP server according to the URL or some predefined rules. In this way, the workload of the HTTP servers can be efficiently balanced.
\par

\paragraph{MQTT load balancers}
There are no functions of load balancing for the MQTT servers present in HAProxy. As a result, TCP load balancing is deployed to distribute load for multiple MQTT servers. TCP load balancing can transfer TCP packages to the matching backend server in accordance with a set of predefined rules. This is realized by a protocol termed Network Address Translation (NAT) and some load balancing methods. According to the NAT protocol, HAProxy provides a virtual IP and port for outer networks to visit, and listen to a certain port for TCP packages. When it receives any TCP package, HAProxy can change the destination IP address and the port number of the package before forwarding the concerned package to the target backend MQTT server. Different from the HTTP servers, the MQTT servers need to keep long-lived connections with clients, and thus HAProxy needs to maintain these connections. Therefore, the Least Connections policy is chosen as the load balancing method for the MQTT servers in HAProxy. It helps HAProxy select the MQTT server with the least number of active connections as the target server.
\par

\subsection{Services of the IoT cloud}
The IoT cloud should meet the requirements of different IoT applications. By using the services provided by the IoT cloud, diverse applications can be offered to both end-users, developers or managers. These services can be accessed by web browsers or smartphones anytime and anywhere. The IoT cloud services can be mainly divided into three categories, i.e.,
\par

\subsubsection{Web applications}
The IoT cloud can provide services through web applications by deploying web pages in HTTP servers. These HTML pages are developed using the Hyper Text Markup Language (HTML) and Cascading Style Sheets (CSS) for static webpages, as well as JavaScript that defines the actions a page should take towards different events. Web applications in the IoT cloud are mainly developed for managers to supervise the devices they own. By managing the IoT cloud interface, managers can monitor the detailed information of the devices, e.g., the on-off state, the Media Access Control (MAC) address or the timing tasks. Besides, they can control devices directly with the permission of device owners for convenient debug.

\subsubsection{Mobile applications}
Smartphone is becoming an indispensable communications tool for people's everyday life. With the pervasiveness of smartphones, people browse the Internet on their phone using a large variety of smartphone application programs (APPs). Android and iOS are the two dominant OSs for smartphones. For the convenience of end-users, mobile APPs based upon the IoT cloud are developed to meet the requirements of both the Android and iOS platforms. On the other hand, there are some APPs that are developed by third parties (such as Facebook or WeChat) that are very popular. In order to provide IoT cloud services to end-users, several interfaces based on these APPs are defined to allow users to access to the corresponding services. For instance, end-users can control their smart-home devices using APPs that are designed by manufacturers or by WeChat directly.
\par

\subsubsection{Software Development Kits}
In order to enhance its applications and services, the IoT cloud provides two Software Development Kits (SDKs) to third party developers, i.e., Android and iOS SDKs. The SDKs consist of several Application Programming Interfaces (APIs) by which the complex functions of the IoT cloud can be easily used. Every API encapsulates a number of underlying operations, and provides easy access for developers. The SDKs make it easier to develop APPs so as to unleash the full potential of the IoT cloud.
\par

\section{Experimental Results and Analysis}
In this Section, our experiments are carried out under various conditions in order to evaluate the performance of the proposed IoT cloud. We focus only on the performance of the MQTT and HTTP servers because both of them have a significant impact on the perceived quality of the services of the IoT cloud under consideration. Because of the different mechanisms of MQTT and HTTP servers, different metrics are used to evaluate their respective performance. 

\subsection{Experimental Configuration}

The MQTT and HTTP servers are deployed on two separate VMs. The number of CPU cores used by either the MQTT or HTTP server can be configured from one to four. For testing purposes, we use Node.js on another VM to simulate the interactions between the clients and servers. Furthermore, a Redis cluster with eight nodes is implemented on four VMs, acting as both the database and broker.

\subsection{Results}

\subsubsection{HTTP server}
To measure the performance of the HTTP server, a certain number of clients is generated, each of which establishes a connection with the HTTP server. The number of clients can be between 1,000 to 20,000 as deemed appropriate. All the clients send only the GET requests to the servers to fetch a web page, whose size is about 1K bytes. If the HTTP server handles the request successfully, it sends the web page back to the client. Once the client receives the response, it continues to send a same new request. Otherwise, if no response is received by the client within ten seconds, the request is regarded as a failure. During a total of 180 seconds, the number of requests that are successfully handled by the HTTP servers is counted so as to compute the throughput performance of the HTTP servers. Meanwhile, the response time of each request is also collected and averaged.
\par

\begin{figure*}[!t]
	\centering
	\subfloat[Average response time of the HTTP server]{
		\includegraphics[width=0.5\textwidth]{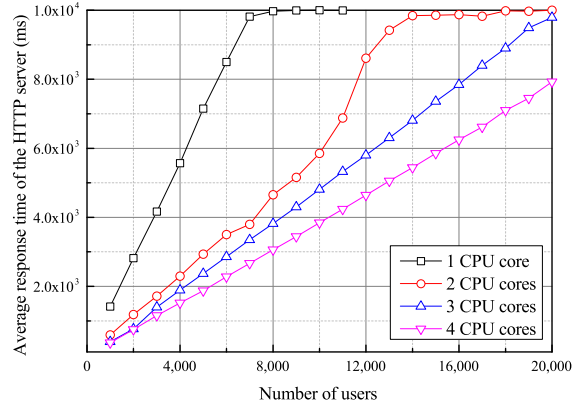}
		\label{fig_http_latency}
	}
	\hfil
	\subfloat[Throughput of the HTTP server]{
		\includegraphics[width=0.5\textwidth]{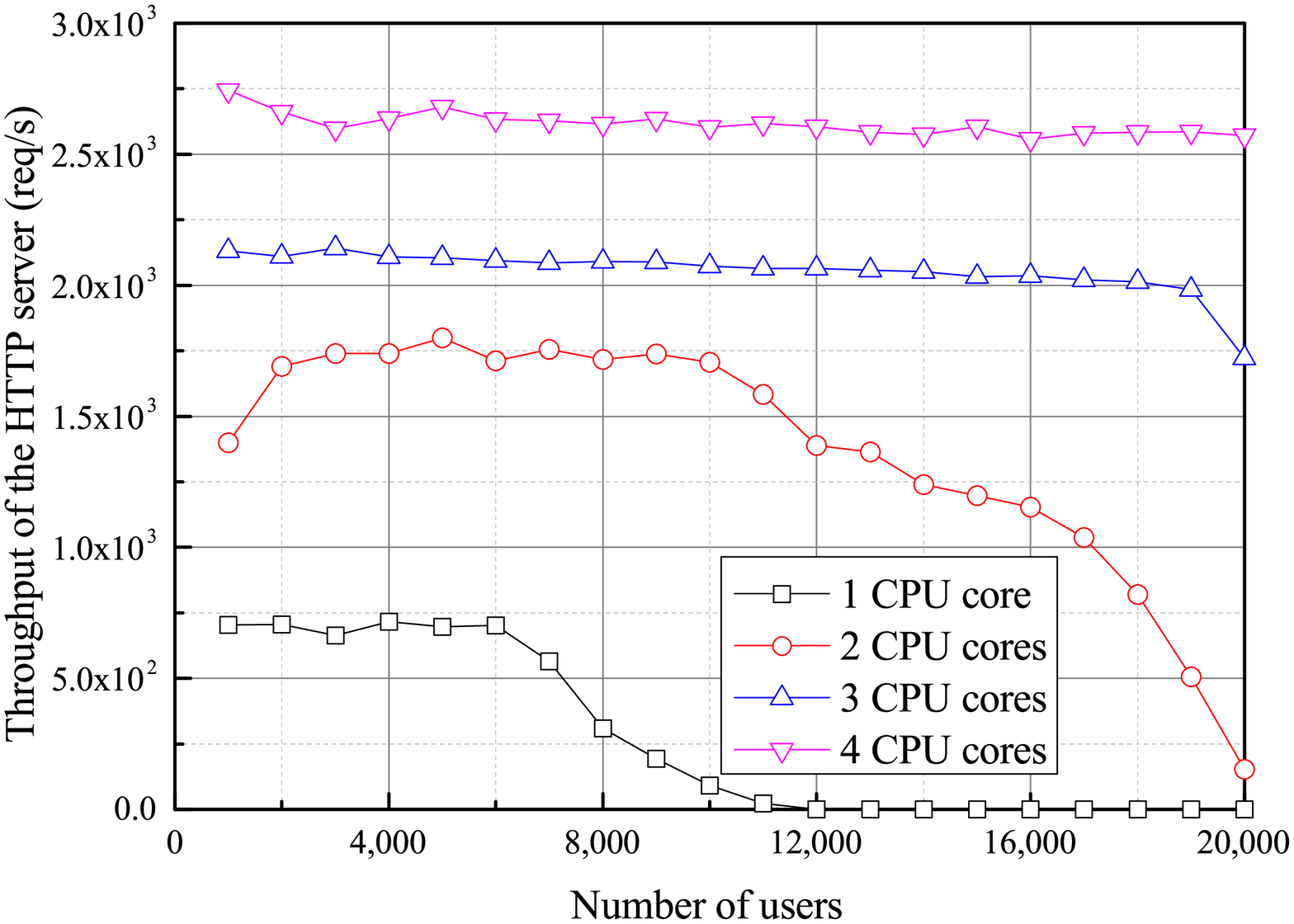}
		\label{fig_http_throughput}
	}
	\subfloat[Average CPU utilization of the HTTP server]{
		\includegraphics[width=0.5\textwidth]{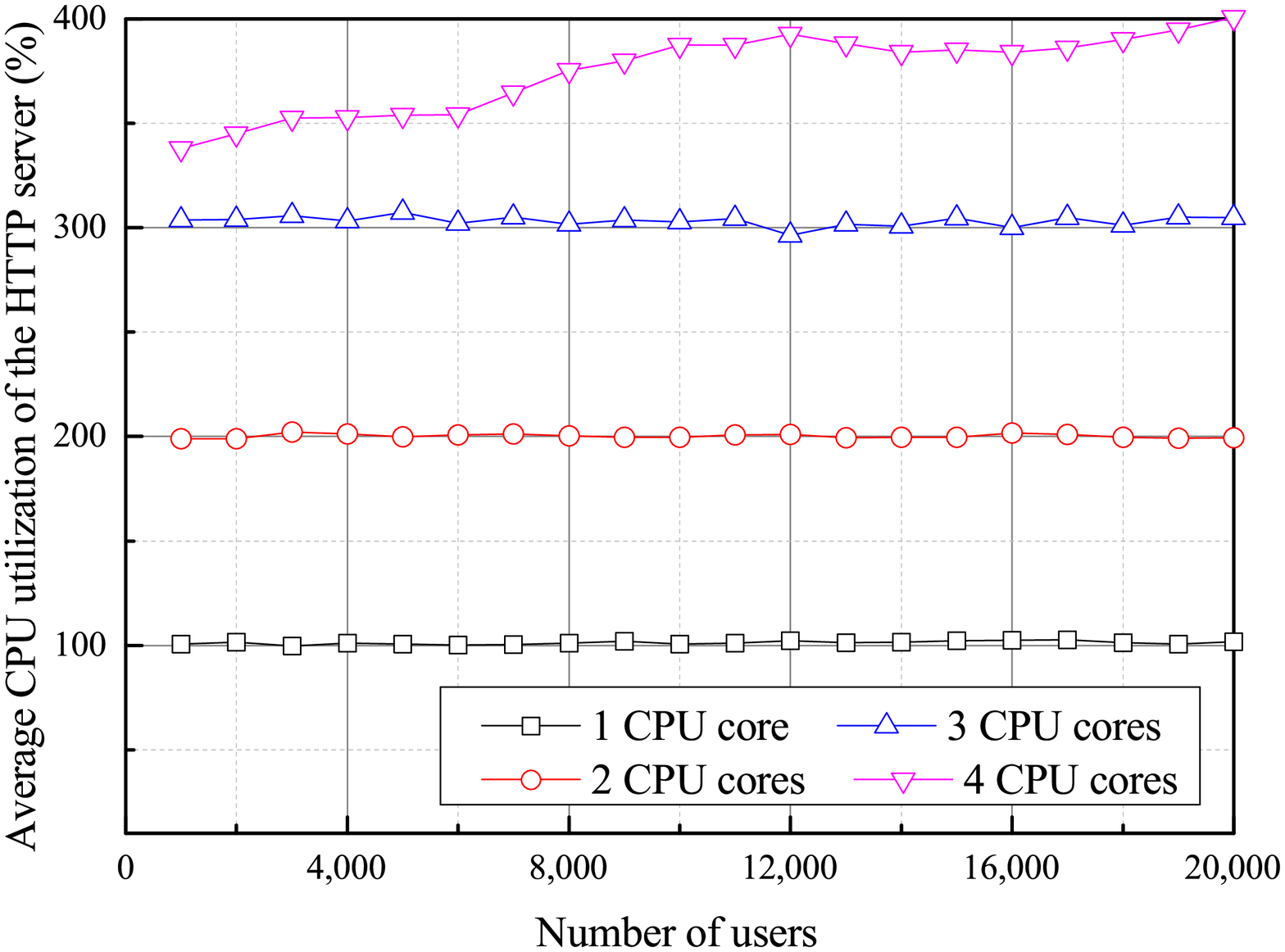}
		\label{fig_http_cpu}
	}
	\caption{Performance of the HTTP server}
	\label{fig_http}
\end{figure*}

Fig.~\subref*{fig_http_latency} shows that the response time becomes largely linear with the number of clients. Since the clients continuously send requests, the HTTP server runs at its full capacity all the time. Thus, throughput and CPU utilization remain at high levels as can be seen from in Fig.~\subref*{fig_http_throughput} and Fig.~\subref*{fig_http_cpu}. However, many requests cannot be handled on time and are discarded when the number of clients is larger than 10,000 or 20,000 for the HTTP server with one or two CPU cores, respectively. Correspondingly, the throughput performance of the HTTP server with one or two CPU cores rapidly deteriorate when the number of clients is above 10,000 or 20,000, respectively. In addition, due to the asynchronism of HTTP server, it can still handle more requests with full CPU utilization as shown in Fig.~\subref*{fig_http_cpu}.
\par

\subsubsection{MQTT server}

We consider 2,000 to 40,000 publishers when evaluating the performance of the MQTT servers. Each publisher subscribes an individual topic and publishes messages that contain only the timestamp to that topic every 10 seconds. During the period of the experiment, i.e., 120 seconds, we collect the information on the total number of messages that have been successfully delivered to the publishers as well as the transmission latency of the messages.
\par

\begin{figure*}[!t]
	\centering
	\subfloat[Average transmission latency of the MQTT server]{
		\includegraphics[width=0.5\textwidth]{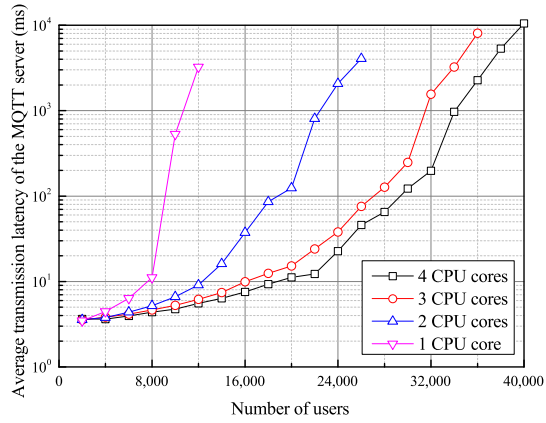}
		\label{fig_mqtt_latency}
		}
		\hfil
	\subfloat[Throughput of the MQTT server]{
		\includegraphics[width=0.5\textwidth]{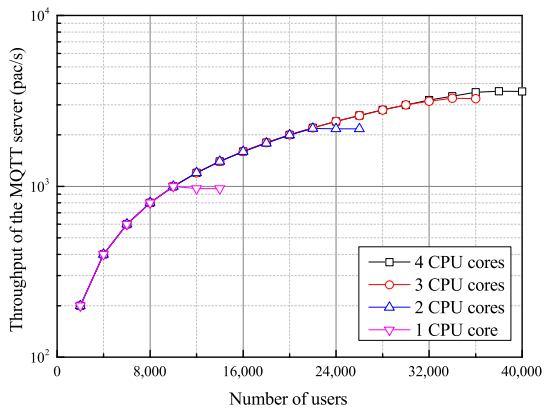}
		\label{fig_mqtt_throughput}
	}
	\subfloat[Average CPU utilization of the MQTT server]{
		\includegraphics[width=0.5\textwidth]{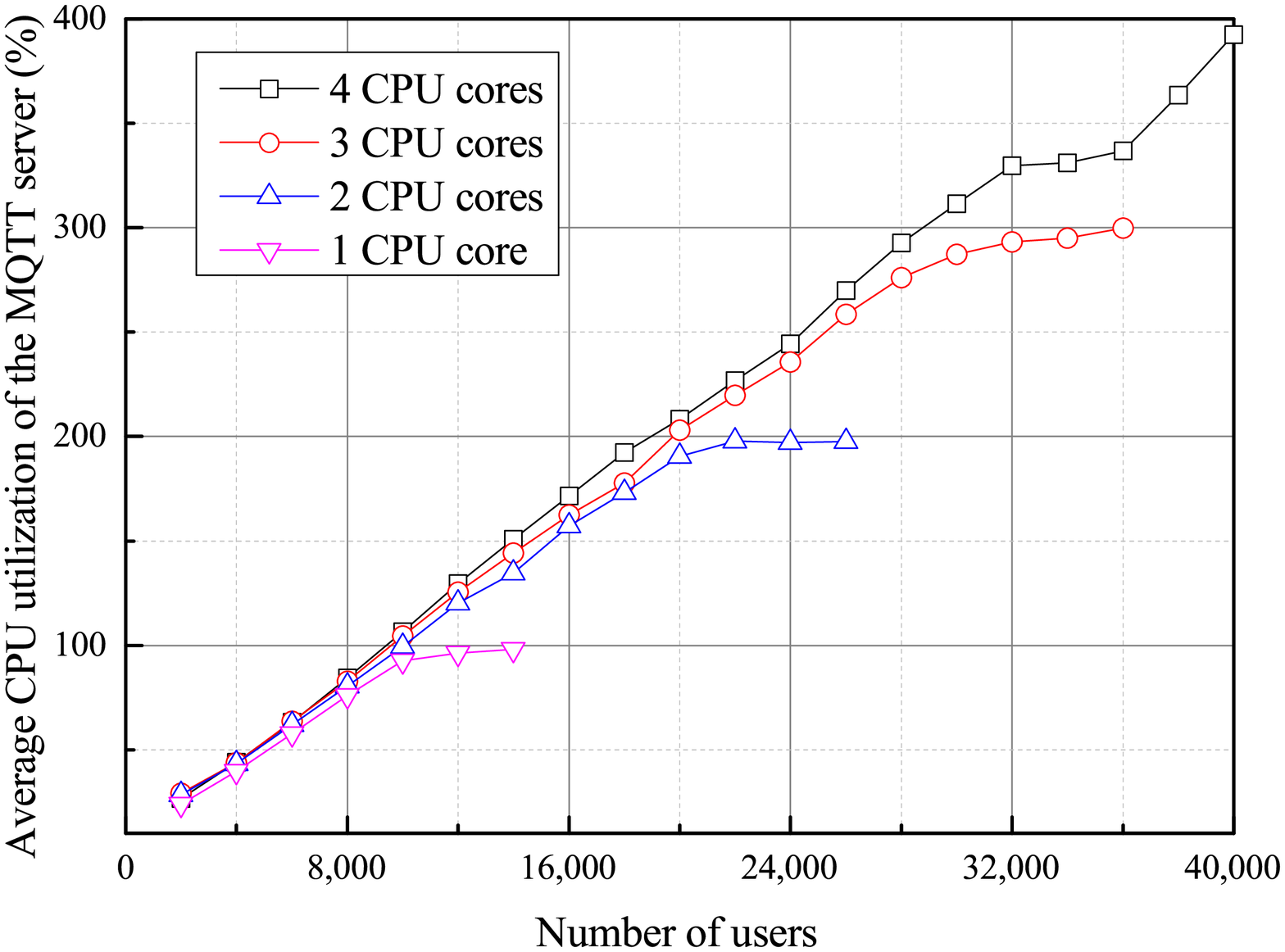}
		\label{fig_mqtt_cpu}
	}
	\caption{Performance of the MQTT server}
	\label{fig_mqtt}
\end{figure*}

As shown in Fig.~\ref{fig_mqtt}, one MQTT server can provide services up to 40,000 clients with four CPU cores in ten-second intervals of publishing at most. The average transmission latency becomes larger with the increase of the number of users, and lower when the number of CPU cores increases. The throughput of the MQTT server becomes floored when the number of clients becomes larger than some predefined threshold, depending on the number of CPU cores, i.e., 10,000, 22,000, 34,000 and 40,000 for one, two, three and four cores, respectively. It is evident that an excessive number of packages render the MQTT server unable to transfer packages in time. A similar trend for CPU utilization can be observed in Fig.~\subref*{fig_mqtt_cpu}.
\par

\section{Conclusion and Outlook}
IoT is a new technological paradigm enabling ubiquitous things or objects to interact with each other and to access to the Internet. By integrating cloud computing and IoT techniques, new valuable and reliable services can be provided to many users. This article mainly proposed an IoT cloud architecture and its corresponding implementation. The key point of the architecture is the combination of the HTTP and MQTT servers, as well as the implementation of the message broker. Several experiments were conducted with the objective of evaluating the performance of the application servers in the proposed IoT cloud. The performance results demonstrated the significant impact of the number of clients and CPU cores on the average transmission latency/response time, throughput and CPU utilization for the HTTP and MQTT servers, respectively. They also study and discuss the performance of the IoT cloud that was implemented. 
\par
Much more research is still needed to address many other challenges of the IoT cloud in the near future. First of all, there is a lack of well-defined standards to unify varying architectures and interfaces of the IoT cloud. Moreover, advanced data analytics are necessary to make full use of big data that IoT brings about. Last but not least, security and privacy challenges should also be taken into account in designing the IoT cloud.
\par

\section*{Biography}

\begin{IEEEbiographynophoto}
	{\bf Lu Hou}
	received his B.S. degree from the School of Information and Communication Engineering, Beijing University of Posts and Telecommunications (BUPT), China, in 2014. He is now a candidate for Ph.D. in the Intelligent Computing and Communication (IC$^2$) Lab, Key Lab of Universal Wireless Communications, Ministry of Education, BUPT.
	Nowadays, he mainly focuses on resource allocation and security in mobile cloud computing.
\end{IEEEbiographynophoto}
\vspace*{-15pt}

\begin{IEEEbiographynophoto}
	{\bf Shaohang Zhao}
	received his BS degree from Beijing University of Posts and Telecommunications (BUPT), China, in 2014. He is now studying for a Master’s Degree in the Intelligent Computing and Communication (IC$^2$) lab. His research mainly concentrates on the Internet of Things networks.
\end{IEEEbiographynophoto}
\vspace*{-15pt}

\begin{IEEEbiographynophoto}
	{\bf Xiong Xiong}
	received his B.S. degree from Beijing University of Posts and Telecommunications (BUPT), China, in 2013. Since then, he has been working toward a Ph.D. degree at BUPT. His research interests include M2M networks and software defined radio.
\end{IEEEbiographynophoto}
\vspace*{-15pt}

\begin{IEEEbiographynophoto}
	{\bf Kan Zheng}
	(S'02-M'06-SM'09) currently a Professor in Beijing University of Posts and Telecommunications (BUPT), China. His current research interests lie in the field of wireless communications, with an emphasis on performance analysis and optimization of heterogeneous networks and 5G networks. He has published more than 200 papers in IEEE conferences and transactions.
\end{IEEEbiographynophoto}
\vspace*{-15pt}

\begin{IEEEbiographynophoto}
	{\bf Periklis Chazimisios}
	(S’02-M’05-SM’12) is currently an Associate Professor with the Computing Systems, Security and Networks (CSSN) Research Lab of the Department of Informatics at the Alexander Technological Educational Institute of Thessaloniki. He currently participates in several European and National Research Projects (such as COST Actions, Arximidis III). 
\end{IEEEbiographynophoto}
\vspace*{-15pt}

\begin{IEEEbiographynophoto}
	{\bf M. Shamim Hossain}
	(SM'09) serves as an Associate Professor and a Division Head for the Department of Informatics at Alexander TEI of Thessaloniki (ATEITHE), Greece. He is involved in several standardization activities and author/editor of 8 books and more than 100 peer-reviewed papers on performance evaluation and standardization of mobile/wireless communications, Internet of Things and Big Data. Dr. Chatzimisios received his Ph.D. from Bournemouth University,UK in 2005 and his B.Sc. from ATEITHE, Greece in 2000. 
\end{IEEEbiographynophoto}
\vspace*{-15pt}

\begin{IEEEbiographynophoto}
	{\bf Wei Xiang}
	(S'00-M'04-SM'10) received the B.Eng. and M.Eng. degrees, both in electronic engineering, from the University of Electronic Science and Technology of China, Chengdu, China, in 1997 and 2000, respectively, and the Ph.D. degree in telecommunications engineering from the University of South Australia, Adelaide, Australia, in 2004. He is currently Foundation Professor and Head of Discipline Electronic Systems and Internet of Things Engineering in the College of Science and Engineering at James Cook University, Cairns, Australia. 
\end{IEEEbiographynophoto}
\vspace*{-15pt}
\enlargethispage{-9.5cm}

\clearpage

\begin{table}
	\centering
	\caption{Features of typical applications using IoT cloud.}
	\label{table_application}
	\begin{tabular}{|c|c|c|c|c|c|c|c|c|c|c|}
		\hline
		\textbf{\makecell*[c]{Scenario}} 			& \textbf{\makecell*[c]{Typical \\use case}} & \textbf{\makecell*[c]{Device \\ type}} & \textbf{\makecell*[c]{User\\ population}} & \textbf{\makecell*[c]{Energy \\ consump-\\tion}} & \textbf{\makecell*[c]{Main-\\tenance \\cost}} & \textbf{\makecell*[c]{Through-\\put}} & \textbf{\makecell*[c]{Tolerable \\ latency}} & \textbf{\makecell*[c]{Mobi-\\lity}} & \textbf{\makecell*[c]{Relia-\\bility}} & \textbf{\makecell*[c]{Security\\ \& \\Privacy}} \\
		\hline
		\multirow{2}{*}{\makecell*[c]{Smart \\ Building}}  & \makecell*[c]{Water\\ metering} & \makecell*[c]{Sensors,\\meters} & \makecell*[c]{Large} &\makecell*[c]{Low} &\makecell*[c]{Low} & \makecell*[c]{Low} & \makecell*[c]{High} & \makecell*[c]{Fixed} & \makecell*[c]{Medium} &\makecell*[c]{Low}\\
		\cline{2-11}
		&\makecell*[c]{Residential\\monitoring}&\makecell*[c]{Sensors}&\makecell*[c]{Few}&\makecell*[c]{Low}&\makecell*[c]{Low}&\makecell*[c]{Low}&\makecell*[c]{Low}&\makecell*[c]{Fixed}&\makecell*[c]{High}&\makecell*[c]{High}\\
		\hline
		\multirow{2}{*}{\makecell*[c]{Smart \\ Home/\\Office}}         & \makecell*[c]{Home\\ automation} & \makecell*[c]{Intelligent\\ appliances, \\ sensors} & \makecell*[c]{Few} &\makecell*[c]{High} &\makecell*[c]{Low} &\makecell*[c]{Low} &\makecell*[c]{High} &\makecell*[c]{Low} &\makecell*[c]{Medium}&\makecell*[c]{High}\\
		\cline{2-11}
		&\makecell*[c]{Smart\\ meeting} & \makecell*[c]{Laptops,\\videos} & \makecell*[c]{Medium} &\makecell*[c]{High} &\makecell*[c]{Low} &\makecell*[c]{High} &\makecell*[c]{Medium} &\makecell*[c]{Fixed} &\makecell*[c]{Low}&\makecell*[c]{High}\\
		\hline
		\multirow{2}{*}{\makecell*[c]{Intelligent \\Transpor-\\tation}} & \makecell*[c]{Traffic \\monitoring} & \makecell*[c]{Sensors,\\cameras} & \makecell*[c]{Large} &\makecell*[c]{High} &\makecell*[c]{Low} &\makecell*[c]{High} &\makecell*[c]{High} &\makecell*[c]{Fixed} &\makecell*[c]{Low}&\makecell*[c]{Low} \\
		\cline{2-11}
		&\makecell*[c]{Driving \\ Assistance} &\makecell*[c]{Sensors, \\vehicles} &\makecell*[c]{Few} &\makecell*[c]{Medium} &\makecell*[c]{Medium} &\makecell*[c]{Low} &\makecell*[c]{Low} &\makecell*[c]{High} &\makecell*[c]{High} &\makecell*[c]{High} \\
		\hline
		\multirow{2}{*}{\makecell*[c]{Smart\\ Healthcare}}   & \makecell*[c]{Patient\\monitoring}  & \makecell*[c]{Sensors,\\medical\\equipment} & \makecell*[c]{Few} &\makecell*[c]{Low} &\makecell*[c]{Low} &\makecell*[c]{Low} &\makecell*[c]{Medium} &\makecell*[c]{Fixed} &\makecell*[c]{High}&\makecell*[c]{High} \\
		\cline{2-11}
		&\makecell*[c]{Vital signal\\alert} & \makecell*[c]{Sensors,\\ medical\\equipment} & \makecell*[c]{Few}  & \makecell*[c]{Low}  &\makecell*[c]{Low}  & \makecell*[c]{Low}  & \makecell*[c]{Low}  &\makecell*[c]{Medium}  & \makecell*[c]{High}  & \makecell*[c]{High}\\
		\hline
	\end{tabular}	
\end{table}

\clearpage

\clearpage

\clearpage

\clearpage

\end{document}